# Generalising tractable VCSPs defined by symmetric tournament pair multimorphisms

Vladimir Kolmogorov*

University College London

v.kolmogorov@cs.ucl.ac.uk

Stanislav Živný†

Oxford University

standa.zivny@comlab.ox.ac.uk

**Abstract**

We study optimisation problems that can be formulated as valued constraint satisfaction problems (VCSP). A problem from VCSP is characterised by a *constraint language*, a fixed set of cost functions taking finite and infinite costs over a finite domain. An instance of the problem is specified by a sum of cost functions from the language and the goal is to minimise the sum. We are interested in *tractable* constraint languages; that is, languages that give rise to VCSP instances solvable in polynomial time. Cohen *et al.* (AIJ'06) have shown that constraint languages that admit the MJN multimorphism are tractable. Moreover, using a minimisation algorithm for submodular functions, Cohen *et al.* (TCS'08) have shown that constraint languages that admit an STP (symmetric tournament pair) multimorphism are tractable.

We generalise these results by showing that languages admitting the MJN multimorphism on a subdomain and an STP multimorphisms on the complement of the subdomain are tractable. The algorithm is a reduction to the algorithm for languages admitting an STP multimorphism.

*Vladimir Kolmogorov is supported by the Royal Academy of Engineering/EPSRC.

†Stanislav Živný is supported by Junior Research Fellowship at University College, Oxford. Part of this work was done while the second author was visiting Microsoft Research Cambridge.

# 1 Introduction

The constraint satisfaction problem is a central generic problem in computer science. It provides a common framework for many theoretical problems as well as for many real-life applications, see [1] for a nice survey. An instance of the *constraint satisfaction problem* (CSP) consists of a collection of variables which must be assigned values subject to specified constraints. CSP is known to be equivalent the problem of evaluating conjunctive queries on databases [2], and to the homomorphism problem for relations structures [3].

An important line of research on the CSP is to identify all tractable cases; that is, cases that are recognisable and solvable in polynomial time. Most of this work has been focused on one of the two general approaches: either identifying structural properties of the way constrains interact which ensure tractability no matter what forms of constraint are imposed [4], or else identifying forms of constraint which are sufficiently restrictive to ensure tractability no matter how they are combined [5, 3].

The first approach has been used to characterise all tractable cases of bounded-arity CSPs: the *only* class of structures which ensures tractability (subject to a certain complexity theory assumption, namely FPT $\neq$ W[1]) are structures of bounded tree-width modulo homomorphic equivalence [6, 7]. The second approach has led to identifying certain algebraic properties known as polymorphisms [8] which are necessary for a set of constraint types to ensure tractability. A set of constraint types which ensures tractability is called a *tractable constraint language*.

Since in practice many constraint satisfaction problems are over-constrained, and hence have no solution, or are under-constrained, and hence have many solutions, *soft* constraint satisfaction problems have been studied [9]. In an instance of the soft CSP, every constraint is associated with a function (rather than a relation as in the CSP) which represents preferences among different partial assignments, and the goal is to find the best assignment. Several very general soft CSP frameworks have been proposed in the literature [10, 11]. In this paper we focus on one of the very general frameworks, the *valued* constraint satisfaction problem (VCSP) [10].

Similarly to the CSP, an important line of research on the VCSP is to identify tractable cases which are recognisable in polynomial time. Is is well known that structural reasons for tractability generalise to the VCSP [9]. In the case of language restrictions, only a few conditions are known to guarantee tractability of a given set of valued constraints [12, 13].

**Related work** Cohen *et al.* have completely classified the complexity of valued languages over Boolean domains [12]. The classification obtained in [12] relies on the notion of a *multimorphism*. In particular, [12] has shown that any language that admits a min-max or the MJN multimorphism[1] is tractable. The min-max multimorphisms have been generalised in [13], where Cohen *et al.* have shown that any language that admits an STP (symmetric tournament pair) multimorphism is tractable.

**Contributions** Using ideas from recent work of the authors [14] and [15], we generalise the tractability results from [12] and [13]. We show that languages admitting the MJN multimorphism on a subdomain and an STP multimorphism on the complement of the subdomain are tractable. We describe an algorithm which, after establishing strong 3-consistency, will reduce the problem to an instance admitting an STP multimorphism.

**Organisation of the paper** The rest of the paper is organised as follows. In Section 2, we define valued constraint satisfaction problems (VCSPs), multimorphisms and other necessary definitions needed throughout the paper. We state our results in Section 3, describe our algorithm in Section 4 and prove the correctness of the algorithm in Section 5.

[1] MJN is defined in Section 2, and called $\langle \mathrm{Mjrty}_1, \mathrm{Mjrty}_2, \mathrm{Mnrty}_3 \rangle$ in [12].

## 2 Background and notation

We denote by $\mathbb{R}_+$ the set of all non-negative real numbers. We denote $\overline{\mathbb{R}}_+ = \mathbb{R}_+ \cup \{\infty\}$ with the standard addition operation extended so that for all $a \in \mathbb{R}_+$, $a + \infty = \infty$. Members of $\overline{\mathbb{R}}_+$ are called *costs*.

Throughout the paper, we denote by $D$ any fixed finite set, called a *domain*. Elements of $D$ are called *domain values* or *labels*.

A function $f$ from $D^m$ to $\overline{\mathbb{R}}_+$ will be called a *cost function* on $D$ of *arity* $m$. If the range of $f$ is $\{0, \infty\}$, then $f$ is called a *crisp* cost function, or just a *relation*. Let $f : D^m \to \overline{\mathbb{R}}_+$ be an $m$-ary cost function $f$. We denote $\mathtt{dom}\, f = \{\boldsymbol{x} \in D^m \mid f(\boldsymbol{x}) < \infty\}$ to be the effective domain of $f$. Functions $f$ of arity $m = 2$ are called *binary*.

A *language* is a set of cost functions with the same set $D$. Language $\Gamma$ is called crisp if all cost functions in $\Gamma$ are crisp.

**Definition 1.** *An instance $\mathcal{I}$ of the* valued constraint satisfaction problem *(VCSP) is a function $D^V \to \overline{\mathbb{R}}_+$ given by*

$$\mathit{Cost}_{\mathcal{I}}(\boldsymbol{x}) = \sum_{t \in T} f_t \left( x_{i(t,1)}, \ldots, x_{i(t,m_t)} \right)$$

*It is specified by a finite set of nodes $V$, finite set of terms $T$, cost functions $f_t : D^{m_t} \to \overline{\mathbb{R}}_+$ or arity $m_t$ and indices $i(t,k) \in V$ for $t \in T$, $k = 1, \ldots, m_t$. A* solution *to $\mathcal{I}$ is an assignment $\boldsymbol{x} \in D^V$ with the minimum cost.*

We denote by $\mathsf{VCSP}(\Gamma)$ the class of all VCSP instances whose terms $f_t$ belong to $\Gamma$. A finite language $\Gamma$ is called *tractable* if $\mathsf{VCSP}(\Gamma)$ can be solved in polynomial time. An infinite language $\Gamma$ is tractable if every finite subset $\Gamma' \subseteq \Gamma$ is tractable.

**Definition 2.** *A mapping $F : D^k \to D$, $k \geq 1$ is called a* polymorphism *of a cost function $f : D^m \to \overline{\mathbb{R}}_+$ if*

$$F(\boldsymbol{x}_1, \ldots, \boldsymbol{x}_k) \in \mathtt{dom}\, f \qquad \forall \boldsymbol{x}_1, \ldots, \boldsymbol{x}_k \in \mathtt{dom}\, f$$

*where $F$ is applied component-wise. $F$ is a polymorphism of a language $\Gamma$ if $F$ is a polymorphism of every cost function in $\Gamma$.*

Multimorphisms [12] are generalisations of polymorphisms. To make the paper easier to read, we only define binary and ternary multimorphisms as we will not need multimorphism of higher arities.

**Definition 3.** *Let $\langle \sqcap, \sqcup \rangle$ be a pair of operations, where $\sqcap, \sqcup : D \times D \to D$, and let $\langle F^1, F^2, F^3 \rangle$ be a triple of operations, where $F^i : D \times D \times D \to D$, $1 \leq i \leq 3$.*

- *Pair $\langle \sqcap, \sqcup \rangle$ is called a (binary)* multimorphism *of cost function $f : D^m \to \overline{\mathbb{R}}_+$ if*

$$f(\boldsymbol{x} \sqcap \boldsymbol{y}) + f(\boldsymbol{x} \sqcup \boldsymbol{y}) \leq f(\boldsymbol{x}) + f(\boldsymbol{y}) \qquad \forall \boldsymbol{x}, \boldsymbol{y} \in \mathtt{dom}\, f \tag{1}$$

  *where operations $\sqcap, \sqcup$ are applied component-wise. $\langle \sqcap, \sqcup \rangle$ is a multimorphism of language $\Gamma$ if $\langle \sqcap, \sqcup \rangle$ is a multimorphism of every $f$ from $\Gamma$.*

- *Triple $\langle F^1, F^2, F^3 \rangle$ is called a (ternary)* multimorphism *of cost function $f : D^m \to \overline{\mathbb{R}}_+$ if*

$$f(F^1(\boldsymbol{x},\boldsymbol{y},\boldsymbol{z})) + f(F^2(\boldsymbol{x},\boldsymbol{y},\boldsymbol{z})) + f(F^3(\boldsymbol{x},\boldsymbol{y},\boldsymbol{z})) \leq f(\boldsymbol{x}) + f(\boldsymbol{y}) + f(\boldsymbol{z}) \quad \forall \boldsymbol{x}, \boldsymbol{y}, \boldsymbol{z} \in \mathtt{dom}\, f \tag{2}$$

  *where operations $F^1, F^2, F^3$ are applied component-wise. $\langle F^1, F^2, F^3 \rangle$ is a multimorphism of language $\Gamma$ if $\langle F^1, F^2, F^3 \rangle$ is a multimorphism of every $f$ from $\Gamma$.*

- *Pair* $\langle \sqcap, \sqcup \rangle$ *is called* conservative *if* $\{a \sqcap b, a \sqcup b\} = \{a, b\}$ *for all* $a, b \in D$. *Operation* $F^i$ *is called* conservative $F^i(a, b, c) \in \{a, b, c\}$ *for all* $a, b, c \in D$.
- *Pair* $\langle \sqcap, \sqcup \rangle$ *is called a* symmetric tournament pair (STP) *if it is conservative and both operations* $\sqcap, \sqcup$ *are commutative, i.e.* $a \sqcap b = b \sqcap a$ *and* $a \sqcup b = b \sqcup a$ *for all* $a, b \in D$.

Theorems 4 and 6 below give two classes of languages that are known to be tractable.

**Theorem 4** ([13]). *If a language* $\Gamma$ *admits an STP multimorphism, then* $\Gamma$ *is tractable.*

**Definition 5.** *Let* $\mathtt{Mj}^1$, $\mathtt{Mj}^2$, *and* $\mathtt{Mn}^3$ *be the ternary functions defined on* $D$ *as follows:*

$$\mathtt{Mj}^1(x,y,z) = \begin{cases} y & \text{if } y = z \\ x & \text{otherwise} \end{cases} \quad \mathtt{Mj}^2(x,y,z) = \begin{cases} x & \text{if } x = z \\ y & \text{otherwise} \end{cases}$$

$$\mathtt{Mn}^3(x,y,z) = \begin{cases} x & \text{if } y = z \text{ and } z \neq x \\ y & \text{if } x = z \text{ and } z \neq y \\ z & \text{otherwise} \end{cases}$$

*We call* $\langle \mathtt{Mj}^1, \mathtt{Mj}^2, \mathtt{Mj}^3 \rangle$ *the MJN multimorphism.*

**Theorem 6** ([12]). *If a language* $\Gamma$ *admits the MJN multimorphism, then* $\Gamma$ *is tractable.*[2]

The class that we will consider will include the two classes above as special cases.

Finally, we define the important notion of expressibility, which captures the idea of introducing auxiliary variables in a VCSP instance and the possibility of minimising over these auxiliary variables. (For crisp languages, this is equivalent to *implementation* [16].)

**Definition 7.** *A cost function* $f : D^m \to \overline{\mathbb{R}}_+$ *is* expressible *over a language* $\Gamma$ *if there exists an instance* $\mathcal{I} \in \textsf{VCSP}(\Gamma)$ *with the set of nodes* $V = \{1, \ldots, m, m+1, \ldots, m+k\}$ *where* $k \geq 0$ *such that*

$$f(\boldsymbol{x}) = \min_{\boldsymbol{y} \in D^k} Cost_{\mathcal{I}}(\boldsymbol{x}, \boldsymbol{y}) \qquad \forall \boldsymbol{x} \in D^m$$

*We define* $\Gamma^*$ *to be the* expressive power *of* $\Gamma$*; that is, the set of all cost functions* $f$ *such that* $f$ *is expressible over* $\Gamma$.

The importance of expressibility is in the following result:

**Theorem 8** ([12]). *For any language* $\Gamma$, $\Gamma$ *is tractable iff* $\Gamma^*$ *is tractable.*

It is easy to observe and well known that every multimorphism of $\Gamma$ is also a multimorphism of $\Gamma^*$ [12]. It follows from the definitions that if $\langle F^1, F^2, F^3 \rangle$ is a multimorphism of a relation $\rho$, then $F^i$ is a polymorphism of $\rho$ for every $1 \leq i \leq 3$.

[2] [12] defines languages with only rational costs, but it is easy to extend the simple algorithm for languages admitting the MJN multimorphisms to languages with real costs.

## 3 Our results

We will generalise the problem slightly: we will allow domains for different variables $i \in V$ to be different; they will be denoted as $D_i$, and we define $\mathcal{D} = \times_{i \in V} D_i$. (This extra flexibility will be very important for describing the algorithm.) As a consequence, operations $\sqcap, \sqcup$ may now act differently on different components of vectors $\boldsymbol{x}, \boldsymbol{y} \in \mathcal{D}$. We denote $\sqcap_i, \sqcup_i : D_i \times D_i \to D_i$ to be the $i$-th operation of $\sqcap, \sqcup$. Similarly, we denote by $F_i^1, F_i^2, F_i^3 : D_i \times D_i \times D_i \to D_i$ to be the $i$-the operation of $\langle F^1, F^2, F^3 \rangle$.

We denote by $P$ the collection of sets $P = (P_i)_{i \in V}$ where $P_i = \{\{a, b\} \mid a, b \in D_i, a \neq b\}$. We denote by $M$ a collection of subsets $M = (M_i)_{i \in V}$, $M_i \subseteq P_i$, and $\overline{M} = (\overline{M}_i)_{i \in V}$, $\overline{M}_i = P_i - M_i$.

**Definition 9.** *Let $\langle \sqcap, \sqcup \rangle$ and $\langle \mathtt{Mj}^1, \mathtt{Mj}^2, \mathtt{Mn}^3 \rangle$ be binary and ternary operations respectively.*

- *Pair $\langle \sqcap, \sqcup \rangle$ is an* STP *on $M$ if for all $i \in V$ pair $\langle \sqcap_i, \sqcup_i \rangle$ is an STP on $M_i$, i.e. $\langle \sqcap_i, \sqcup_i \rangle$ is conservative on $P_i$ and commutative on $M_i$.*
- *Triple $\langle \mathtt{Mj}^1, \mathtt{Mj}^2, \mathtt{Mn}^3 \rangle$ is an* MJN *on $\overline{M}$ if for all $i \in V$ operations $\mathtt{Mj}_i^1, \mathtt{Mj}_i^2, \mathtt{Mn}_i^3$ are conservative and for each triple $(a, b, c) \in D_i^3$ with $\{a, b, c\} = \{x, y\} \in \overline{M}_i$ operations $\mathtt{Mj}_i^1(a, b, c)$, $\mathtt{Mj}_i^2(a, b, c)$ return the majority element among $a, b, c$ (that occurs twice) and $\mathtt{Mn}_i^3(a, b, c)$ returns the remaining minority element.*

Our main result is the following

**Theorem 10.** *If $\Gamma$ admits an STP on $M$ and an MJN on $\overline{M}$, for some $M = (M_i)_{i \in V}$ with $M_i \subseteq P_i = \{\{a, b\} \mid a, b \in D_i, a \neq b\}$, then $\Gamma$ is tractable.*

## 4 Algorithm

The idea for the algorithm and some of the proof techniques have been influenced by the techniques used by Takhanov [15] for proving the absence of *arithmetical deadlocks* in certain instances.

We will assume without loss of generality that $\langle \sqcap_i, \sqcup_i \rangle$ is non-commutative on $\{a, b\} \in \overline{M}_i$ (if not, we can simply add such $\{a, b\}$ to $M_i$).

### Stage 1: Decomposition into binary relations

Since the instance admits a majority polymorphism (see Section 5), every cost function $f$ can be decomposed [17] into unary relations $\rho_i \subseteq D_i$, $i \in D_i$ and binary relations $\rho_{ij} \subseteq D_i \times D_j$, $i, j \in V$, $i \neq j$ such that

$$\boldsymbol{x} \in \operatorname{dom} f \qquad \Leftrightarrow \qquad [x_i \in \rho_i \ \forall i \in V] \quad \text{and} \quad [(x_i, x_j) \in \rho_{ij} \ \forall i, j \in V, i \neq j]$$

We will always assume that binary relations are symmetric, i.e. $(x, y) \in \rho_{ij} \Leftrightarrow (y, x) \in \rho_{ji}$. We use the following notation for relations:

- If $\rho_{ij} \in D_i \times D_j$, $X \subseteq D_i$ and $Y \subseteq D_j$ then

  $$\rho_{ij}(X, \cdot) = \{y \mid \exists x \in X \text{ s.t. } (x, y) \in \rho_{ij}\} \qquad \rho_{ij}(\cdot, Y) = \{x \mid \exists y \in Y \text{ s.t. } (x, y) \in \rho_{ij}\}$$

  If $X = \{x\}$ and $Y = \{y\}$ then these two sets will be denoted as $\rho_{ij}(x, \cdot)$ and $\rho_{ij}(\cdot, y)$ respectively.
- If $\rho \in D_1 \times D_2$ and $\rho' \in D_2 \times D_3$ then we define their composition as

  $$\rho \circ \rho' = \{(x, z) \in D_1 \times D_3 \mid \exists y \in D_2 \text{ s.t. } (x, y) \in \rho, (y, z) \in \rho'\}$$

In the first stage we establish *strong 3-consistency* using the standard constraint-processing techniques [18] so that the resulting relations satisfy

$$\begin{array}{llcll} \text{(arc-consistency)} & \{x \mid (x,y) \in \rho_{ij}\} & = & \rho_i & \forall \text{ distinct } i,j \in V \\ \text{(path-consistency)} & \rho_{ik}(x,\cdot) \cap \rho_{jk}(y,\cdot) & \neq & \varnothing & \forall \text{ distinct } i,j,k \in V, (x,y) \in \rho_{ij} \end{array}$$

It is known that in the presence of a majority polymorphism strong 3-consistency is equivalent to global consistency [19]; that is $\mathtt{dom} f$ is empty iff all $\rho_i$ and $\rho_{ij}$ are empty. Using this fact, it is not difficult to show that the strong 3-consistency relations $\rho_i$, $\rho_{ij}$ are uniquely determined by $f$ via

$$\rho_i = \{x_i \mid \boldsymbol{x} \in \mathtt{dom} f\} \qquad \rho_{ij} = \{(x_i, x_j) \mid \boldsymbol{x} \in \mathtt{dom} f\}$$

The second equation implies that any polymorphism of $f$ is also a polymorphism of $\rho_{ij}$.

From now on we will assume that $D_i = \rho_i$ for all $i \in V$. This can be achieved by reducing sets $D_i$ if necessary. We will also assume that all sets $D_i$ are non-empty.

## Stage 2: Modifying $M$ and $\langle \sqcap, \sqcup \rangle$

The second stage of the algorithm works by iteratively growing sets $M_i$ and simultaneously modifying operations $\langle \sqcap_i, \sqcup_i \rangle$ so that (i) $\langle \sqcap_i, \sqcup_i \rangle$ is still a conservative pair which is commutative on $M_i$ and non-commutative on $\overline{M}_i$, and (ii) $\langle \sqcap, \sqcup \rangle$ is a multimorphism of $f$. It stops when we get $M_i = P_i$ for all $i \in V$.

We now describe one iteration. First, we identify subset $U \subseteq V$ and subsets $A_i, B_i \subseteq D_i$ for each $i \in U$ using the following algorithm:

---

1: pick node $k \in V$ and pair $\{a, b\} \in \overline{M}_k$. (If they do not exist, terminate and go to Stage 3.)
2: set $U = \{k\}$, $A_k = \{a\}$, $B_k = \{b\}$
3: **while** there exists $i \in V - U$ such that $\rho_{ki}(A_k, \cdot) \cap \rho_{ki}(B_k, \cdot) = \varnothing$ **do**
4: add $i$ to $U$, set $A_i = \rho_{ki}(A_k, \cdot)$, $B_i = \rho_{ki}(B_k, \cdot)$
// *compute closure of sets* $A_i$ *for* $i \in U$
5: **while** there exists $a \in D_k - A_k$ s.t. $a \in \rho_{ki}(\cdot, A_i)$ for some $i \in U - \{k\}$ **do**
6: add $a$ to $A_k$, set $A_j = \rho_{kj}(A_k, \cdot)$ for all $j \in U - \{k\}$
7: **end while**
// *compute closure of sets* $B_i$ *for* $i \in U$
8: **while** there exists $b \in D_k - B_k$ s.t. $b \in \rho_{ki}(\cdot, B_i)$ for some $i \in U - \{k\}$ **do**
9: add $b$ to $B_k$, set $B_j = \rho_{kj}(B_k, \cdot)$ for all $j \in U - \{k\}$
10: **end while**
// *done*
11: **end while**
12: **return** set $U \subseteq V$ and sets $A_i, B_i \subseteq D_i$ for $i \in U$

---

**Lemma 11.** *Sets* $U$ *and* $A_i, B_i$ *for* $i \in U$ *produced by the algorithm have the following properties:*

*(a)* *Sets* $A_i$ *and* $B_i$ *for* $i \in U$ *are disjoint.*

*(b)* $\{a, b\} \in \overline{M}_i$ *for all* $i \in U$, $a \in A_i$, $b \in B_i$.

*(c)* $\rho_{ki}(A_k, \cdot) = A_i$, $\rho_{ki}(B_k, \cdot) = B_i$, $\rho_{ki}(\cdot, A_i) = A_k$, $\rho_{ki}(\cdot, B_i) = B_k$ *for all* $i \in U - \{k\}$ *where* $k$ *is the node chosen in line 1.*

(d) *Suppose that* $i \in U$ *and* $j \in \overline{U} \equiv V - U$. *If* $(c, x) \in \rho_{ij}$ *where* $c \in A_i \cup B_i$ *and* $x \in D_j$ *then* $(d, x) \in \rho_{ij}$ *for all* $d \in A_i \cup B_i$.

To complete the iteration, we modify sets $M_i$ and operations $\sqcap_i, \sqcup_i$ for each $i \in U$ as follows:

- add all pairs $\{a, b\}$ to $M_i$ where $a \in A_i$, $b \in B_i$.
- redefine $a \sqcap_i b = b \sqcap_i a = a$, $a \sqcup_i b = b \sqcup_i a = b$ for $a \in A_i$, $b \in B_i$

**Lemma 12.** *The new pair of operations* $\langle \sqcap, \sqcup \rangle$ *is a multimorphism of* $f$.

A proof of Lemmas 11 and 12 is given in the next section. They imply that all steps are well-defined, and upon termination the algorithm produces a pair $\langle \sqcap, \sqcup \rangle$ which is an STP multimorphism of $f$.

### Stage 3: Reduction to a submodular minimisation problem

At this stage we have an STP multimorphism. Hence, the instance can be solved by Theorem 4.

## 5 Algorithm's correctness

First, we show that $f$ admits a majority polymorphism $\mu$ using the argument from [15]. Define

$$\begin{aligned} \bar{\mu}(\boldsymbol{x}, \boldsymbol{y}, \boldsymbol{z}) &= [(\boldsymbol{y} \sqcup \boldsymbol{x}) \sqcap (\boldsymbol{y} \sqcup \boldsymbol{z})] \sqcap (\boldsymbol{x} \sqcup \boldsymbol{z}) \\ \mu(\boldsymbol{x}, \boldsymbol{y}, \boldsymbol{z}) &= \texttt{Mj}^1(\bar{\mu}(\boldsymbol{x}, \boldsymbol{y}, \boldsymbol{z}), \bar{\mu}(\boldsymbol{y}, \boldsymbol{z}, \boldsymbol{x}), \bar{\mu}(\boldsymbol{z}, \boldsymbol{x}, \boldsymbol{y})) \end{aligned}$$

Suppose that $\{x, y, z\} = \{a, b\} \in P_i$. It can be checked that $\bar{\mu}_i(x, y, z)$ acts as the majority operation if $\{a, b\} \in M_i$, and $\bar{\mu}_i(x, y, z) = x$ if $\{a, b\} \in \overline{M}_i$. This implies that $\mu_i$ acts as the majority operation on $P_i$.

**Proposition 13.** *If* $\{a, b\} \in \overline{M}_i$, $\{a', b'\} \in P_j$ *and* $(a, a'), (b, b') \in \rho_{ij}$, *where* $i, j$ *are distinct nodes in* $V$, *then exactly one of the following holds:*

(i) $(a, b'), (b, a') \in \rho_{ij}$

(ii) $(a, b'), (b, a') \notin \rho_{ij}$ *and* $\{a', b'\} \in \overline{M}_j$

*Proof.* First, suppose that $\{a', b'\} \in M_j$. We need to show that case (i) holds. Operations $\sqcap_i, \sqcup_i$ are non-commutative on $\{a, b\}$, while $\sqcap_j, \sqcup_j$ are commutative on $\{a', b'\}$. It is easy to check that

$$\{(a, b) \sqcap (a', b'), (a', b') \sqcap (a, b), (a, b) \sqcup (a', b'), (a', b') \sqcup (a, b)\} = \{(a, a'), (a, b'), (a', b), (a', b')\}$$

Since $\sqcap$, $\sqcup$ are polymorphisms of $\rho_{ij}$, all assignments involved in the equation above belong to $\rho_{ij}$. Thus, (i) holds.

Now suppose $\{a', b'\} \in \overline{M}_j$. We then have

$$\texttt{Mn}^3((a, a'), (b, b'), (a, b')) = (b, a') \qquad \texttt{Mn}^3((a, a'), (b, b'), (b, a')) = (a, b')$$

$\texttt{Mn}^3$ is a polymorphism of $\rho_{ij}$, therefore if one of the assignments $(a, b'), (b, a')$ belongs to $\rho_{ij}$ then the other one also belongs to $\rho_{ij}$. This proves the proposition. ∎

### 5.1 Proof of Lemma 11(a-c)

It follows from construction that during all stages of the algorithm there holds

$$\rho_{ki}(A_k,\cdot)=A_i\ ,\quad \rho_{ki}(B_k,\cdot)=B_i\qquad\qquad \forall i\in U-\{k\}\tag{3}$$

Strong 3-consistency also implies that sets $A_i, B_i$ for $i \in U$ are non-empty. Clearly, properties (a) and (b) of Lemma 11 hold after initialization (line 2). Let us prove that each step of the algorithm preserves these two properties. Note, property (a) together with (3) imply that $(a, b') \notin \rho_{ki}$ if $a \in A_k$, $b' \in B_i$, and $(b, a') \notin \rho_{ki}$ if $b \in B_k$, $a' \in A_i$, where $i \in U - \{k\}$.

First, consider line 4, i.e. adding $i$ to $U$ with $A_i = \rho_{ki}(A_k,\cdot)$, $B_i = \rho_{ki}(B_k,\cdot)$. Property (a) for node $i$ follows from the precondition of line 3; let us show (b) for node $i$. Suppose that $a' \in A_i$, $b' \in B_i$, then there exist $a \in A_k$, $b \in B_k$ such that $(a, a'), (b, b') \in \rho_{ki}$. We have $(a, b') \notin \rho_{ki}$, so by Proposition 13 we get $\{a', b'\} \in \overline{M}$.

Now consider line 6, i.e. adding $a$ to $A_k$ and updating $A_j$ for $j \in U - \{k\}$ accordingly. We denote $A^\circ_j$ and $A_j$ to be respectively the old and the new set for node $j \in U$. There must exist node $i \in U - \{k\}$ and element $a' \in A^\circ_i$ such that $(a, a') \in \rho_{ki}$. We prove below that properties (a) and (b) are preserved for nodes $k$, $i$ and all nodes $j \in U - \{k, i\}$.

**<u>Node $k$</u>** It is clear that $a \notin B_k$, otherwise we would have $a' \in \rho_{ki}(B_k,\cdot) = B_i$ contradicting to condition $A^\circ_i \cap B_i = \varnothing$. Thus, property (a) for node $k$ holds. Consider element $b \in B_k$. By arc-consistency there exists element $b' \in \rho_{ki}(b,\cdot) \subseteq B_i$. From property (b) we get $\{a', b'\} \in \overline{M}_i$. We also have $(b, a') \notin \rho_{ki}$ since $A^\circ_i \cap \rho_{ki}(B_k,\cdot) = A^\circ_i \cap B_i = \varnothing$. By Proposition 13 we get $\{a, b\} \in \overline{M}_k$. Thus, property (b) holds for node $k$.

**<u>Node $i$</u>** Let us prove that $A_i \cap B_i = \varnothing$. Suppose not, then $(a, b') \in \rho_{ki}$ for some $b' \in B_i$. There must exist $b \in B_k$ with $(b, b') \in \rho_{ki}$. We have $\rho_{ki} \cap (\{a, b\} \times \{a', b'\}) = \{(a, a'), (b, b'), (a, b')\}$ and $\{a', b'\} \in \overline{M}_i$, which is a contradiction by Proposition 13. This proves property (a) for node $i$.

Property (b) for node $i$ follows from property (a) for nodes $k, i$, property (b) for node $k$, and Proposition 13.

**<u>Node $j \in U - \{k, i\}$</u>** Let us prove that $A_j \cap B_j = \varnothing$. Suppose not, then $(a, y) \in \rho_{kj}$ for some $y \in B_j$. There must exist $b \in B_k$ with $(b, y) \in \rho_{kj}$, and $b' \in B_i$ with $(b, b') \in \rho_{ki}$. We also have $a' \in A^\circ_i = \rho_{ki}(A^\circ_k,\cdot)$, therefore there must exist $c \in A^\circ_k$ with $(c, a') \in \rho_{ki}$, and $x \in A^\circ_{kj}$ with $(c, x) \in \rho_{kj}$. It can be seen that

$$\rho_{ki}\cap\big(\{a,c,b\}\times\{a',b'\}\big)=\{(a,a'),(c,a'),(b,b')\}\qquad \rho_{kj}\cap(\{a,c,b\}\times\{x,y\})=\{(a,y),(c,x),(b,y)\}$$

Indeed, all listed assignments belong to $\rho_{ki}$ or $\rho_{kj}$ by construction; we need to show that remaining assignments do not belong to these relations. We have $(a, b'), (c, b'), (b, a') \notin \rho_{ki}$ since we have already established property (a) for nodes $k$ and $i$. We also have $(c, y), (b, x) \notin \rho_{kj}$ since $A^\circ_k \cap B_k = \varnothing$ and $A^\circ_j \cap B_j = \varnothing$. Combining it with the fact that $\{x, y\} \in \overline{M}$ and using Proposition 13 gives that $(a, x) \notin \rho_{kj}$.

Consider relation $\beta_{ij} = \rho'_{ik} \circ \rho_{kj}$ where $\rho'_{ik} = \{(d', d) \in \rho_{ik} \mid d \in \{a, b, c\}\}$. It is easy to check that $(a', x), (a', y), (b', y) \in \beta_{ij}$ and $(b', x) \notin \beta_{ij}$. We have $\{a', b'\} \in \overline{M}_i$ and $\{x, y\} \in \overline{M}_j$, so $\texttt{Mn}^3((a', x), (a', y), (b', y)) = (b', x)$. Clearly, $\texttt{Mn}^3$ is a polymorphism of $\rho'_{ik}$ and $\beta_{ij}$, therefore we must have $(b', x) \in \beta_{ij}$ - a contradiction. This proves property (a) for node $j$.

Property (b) for node $j$ follows from property (a) for nodes $k, j$, property (b) for node $k$, and Proposition 13.

**Concluding remark** We showed that throughout the algorithm sets $U, A_i, B_i$ satisfy properties (a,b) and equation (3). It is easy to see that after running lines 5-7 we also have $\rho_{ki}(\cdot, A_i) = A_k$, and after running lines 8-10 we have $\rho_{ki}(\cdot, B_i) = B_k$. Thus, property (c) holds upon termination, which concludes the proof of Lemma 11.

## 5.2 Proof of Lemma 11(d)

First, we will prove the following claim:

**Proposition 14.** *Suppose that* $(a, x), (b, x), (c, y) \in \rho_{ij}$ *where* $i \in U$, $j \in \overline{U}$, $a \in A_i$, $b \in B_i$, $c \in A_i \cup B_i$, $x, y \in D_j$. *Then* $(a, y), (b, y), (c, x) \in \rho_{ij}$.

*Proof.* We claim that there exists a relation $\gamma_i \subseteq D_i \times D_i$ with the following properties:

(i) $\gamma_i$ is an equivalence relation, i.e. there exists a unique partitioning $\pi[\gamma_i] = \{C_1, \ldots, C_p\}$ of $D_i$ such that $(x, y) \in \gamma_i$ for $x, y \in D_i$ iff $x$ and $y$ belong to the same partition of $\pi[\gamma_i]$;

(ii) $A_i \in \pi[\gamma_i]$ and $B_i \in \pi[\gamma_i]$;

(iii) operation $\mathtt{Mn}_i^3$ is a polymorphism of $\gamma_i$.

Indeed, for $i = k$ such relation can be constructed as follows. Let us set $\gamma_k = \{(a, a) \mid a \in D_k\}$ and iteratively update it via $\gamma_k := \gamma_k \circ \rho_{ki} \circ \rho_{ik}$ for $i \in U - \{k\}$. Set $\gamma_i$ will never shrink; we stop when no such operation can change $\gamma_k$. Clearly, at this point $\gamma_i$ is an equivalence relation. By comparing this scheme with lines 5-10 of the algorithm we conclude that (ii) holds. Finally, (iii) follows from the fact that polymorphisms are preserved under compositions. If $i \in U - \{k\}$ then we take $\gamma_i = \rho_{ik} \circ \gamma_k \rho_{ki}$; (i)-(iii) then follow from property (c) of Lemma 11.

We are now ready to prove Proposition 14. We can assume that $x \neq y$, otherwise the claim is trivial. Assume that $c \in A_i$ (the case $c \in B_i$ is analogous). Suppose that $(b, y) \notin \rho_{ij}$. We have $\{b, c\} \in \overline{M}$, so Proposition 13 implies that $\{x, y\} \in \overline{M}$. Consider relation $\gamma_i' = \{(x, y) \in \gamma_i \mid y \notin B_i - \{b\})\}$. Polymorphisms in property (iii) are conservative, therefore they are polymorphisms of $\gamma_i'$ as well. Define relation $\beta_{ij} = \gamma_i' \circ \rho_{ij} \subseteq D_i \times D_j$, then $\mathtt{Mn}^3$ is a polymorphism of $\beta_{ij}$. It is easy to check that $(a, y), (a, x), (b, x) \in \beta_{ij}$. Operation $\mathtt{Mn}^3$ is a polymorphism of $\beta_{ij}$ and it acts as the minority operation on $\{a, b\} \in \overline{M}$ and $\{x, y\} \in \overline{M}$, therefore $\mathtt{Mn}^3((a, y), (a, x), (b, x)) = (b, y) \in \beta_{ij}$. This implies that $(b, y) \in \rho_{ij}$, contradicting to the assumption made earlier. We showed that we must have $(b, y) \in \rho_{ij}$. The fact that $\{a, b\} \in \overline{M}$ and Proposition 13 then imply that $(a, y) \in \rho_{ij}$. Finally, the fact that $\{c, b\} \in \overline{M}$ and Proposition 13 imply that $(c, x) \in \rho_{ij}$. Proposition 14 is proved. □

We can now prove Lemma 11(d) under the following assumption:

($*$) Sets $\rho_{ij}(A_i, \cdot)$ and $\rho_{ij}(B_i, \cdot)$ have non-empty intersection.

(This assumption clearly holds if $i = k$, otherwise the algorithm wouldn't have terminated; we will later show that ($*$) holds for nodes $i \in U - \{k\}$ as well.)

First, let us prove that $\rho_{ij}(A_i, \cdot) = \rho_{ij}(B_i, \cdot)$. Suppose that $y \in \rho_{ij}(A_i, \cdot)$, then $(c, y) \in \rho_{ij}$ for some $c \in A_i$. From (d′) we get that there exist $a \in A_i$, $b \in B_i$, $x \in D_j$ such that $(a, x), (b, x) \in \rho_{ij}$. Proposition 14 implies that $(b, y) \in \rho_{ij}$, and thus $\rho_{ij}(A_i, \cdot) \subseteq \rho_{ij}(B_i, \cdot)$. By symmetry we also have $\rho_{ij}(B_i, \cdot) \subseteq \rho_{ij}(A_i, \cdot)$, implying $\rho_{ij}(A_i, \cdot) = \rho_{ij}(B_i, \cdot)$.

Second, let us prove that if $(a, x) \in \rho_{ij}$ where $a \in A_i$, $x \in D_j$ then $(c, x) \in \rho_{ij}$ for all $c \in B_i$. (We call this claim [AB]). As we showed in the previous paragraph, there exists $b \in B_i$ such that $(b, x) \in \rho_{ij}$. We can also select $y \in D_j$ such that $(c, y) \in \rho_{ij}$. Proposition 14 implies that $(c, x) \in \rho_{ij}$, as desired.

A symmetrical argument shows that if $(b, x) \in \rho_{ij}$ where $b \in B_i$, $x \in D_j$ then $(c, x) \in \rho_{ij}$ for all $c \in A_i$ [BA]. By combining facts [AB] and [BA] we obtain that if $(a, x) \in \rho_{ij}$ where $a \in A_i$, $x \in D_j$ then $(c, x) \in \rho_{ij}$ for all $c \in A_i$ [AA], and also that if $(b, x) \in \rho_{ij}$ where $b \in B_i$, $x \in D_j$ then $(c, x) \in \rho_{ij}$ for all $c \in B_i$ [BB].

We have proven Lemma 11(d) assuming that ($*$) holds (and in particular, for $i = k$). It remains to show that ($*$) holds for $i \in U - \{k\}$. Let us select $(a', x) \in \rho_{ij}$ where $a' \in A_i$, $x, y \in D_j$. By strong 3-consistency there exists $a \in D_k$ such that $(a, a') \in \rho_{ki}$ and $(a, x) \in \rho_{kj}$. By Lemma 11(c) we get that $a \in A_k$. As we have just shown, there exists $b \in B_k$ such that $(b, x) \in \rho_{kj}$. By strong 3-consistency there exists $b' \in D_i$ such that $(b, b') \in \rho_{ki}$ and $(b', x) \in \rho_{ij}$. By Lemma 11(c) we get that $b' \in B_i$. We have shown that $x \in \rho_{ij}(A_i, \cdot)$ and $x \in \rho_{ij}(B_i, \cdot)$, which proves (∗).

### 5.3 Proof of Lemma 12

Suppose we have an arc- and path-consistent instance with an STP on $M$ and MJN on $\overline{M}$ and non-empty subset $U$ with $A_i, B_i \subsetneq D_i$ for $i \in U$ that satisfy properties (a-d) of Lemma 11 (where node $k \in U$ is fixed). Let us denote $M^\circ$ and $M$ to be the set before and after the update respectively. Similarly, $\langle \sqcap^\circ, \sqcup^\circ \rangle$ and $\langle \sqcap, \sqcup \rangle$ denote operations before and after the update. We need to show that

$$f(\boldsymbol{x} \sqcap \boldsymbol{y}) + f(\boldsymbol{x} \sqcup \boldsymbol{y}) \leq f(\boldsymbol{x}) + f(\boldsymbol{y}) \qquad \text{if } \boldsymbol{x}, \boldsymbol{y} \in \texttt{dom} f \tag{4}$$

For a vector $\boldsymbol{z} \in \mathcal{D}$ and subset $S \subseteq V$ we denote $\boldsymbol{z}^S$ to be the restriction of $\boldsymbol{z}$ to $S$. Given $\boldsymbol{x}, \boldsymbol{y} \in \mathcal{D}$, denote

$$\delta(\boldsymbol{x}, \boldsymbol{y}) = \begin{cases} 0 & \text{if } \boldsymbol{x}^U \sqcap \boldsymbol{y}^U = \boldsymbol{x}^U \sqcap^\circ \boldsymbol{y}^U \\ 1 & \text{otherwise} \end{cases} \qquad \Delta(\boldsymbol{x}, \boldsymbol{y}) = \{ i \in \overline{U} \mid x_i \neq y_i \}$$

Note, if $\delta(\boldsymbol{x}, \boldsymbol{y}) = 0$ then $\boldsymbol{x} \sqcap \boldsymbol{y} = \boldsymbol{x} \sqcap^\circ \boldsymbol{y}$ and $\boldsymbol{x} \sqcup \boldsymbol{y} = \boldsymbol{x} \sqcup^\circ \boldsymbol{y}$, so the claim is trivial. Let us introduce a total order $\preceq$ on pairs $(\boldsymbol{x}, \boldsymbol{y})$ as the lexicographical order on vector $(|\Delta(\boldsymbol{x}, \boldsymbol{y})|, \delta(\boldsymbol{x}, \boldsymbol{y}))$ (the first component is more significant than the second). We use induction on this order. The base of the induction follows from the following lemma.

**Lemma 15.** *Condition (4) holds for all $\boldsymbol{x}, \boldsymbol{y} \in \texttt{dom} f$ with $|\Delta(\boldsymbol{x}, \boldsymbol{y})| \leq 1$.*

*Proof.* We can assume that $\delta(\boldsymbol{x}, \boldsymbol{y}) = 1$, otherwise the claim holds trivially. Thus, there exists node $i \in U$ such that either $x_i \in A_i, y_i \in B_i$ or $x_i \in B_i, y_i \in A_i$, Lemma 11(c) implies that either $x_i \in A_i, y_i \in B_i$ for all $i \in U$ or $x_i \in B_i, y_i \in A_i$ for all $i \in U$. Therefore, from the definition of operations $\sqcap, \sqcup$ we get $\{\boldsymbol{x}^U \sqcap \boldsymbol{y}^U, \boldsymbol{x}^U \sqcup \boldsymbol{y}^U\} = \{\boldsymbol{x}^U, \boldsymbol{y}^U\}$. Also, we have $\boldsymbol{x} \sqcap^\circ \boldsymbol{y}, \boldsymbol{x} \sqcup^\circ \boldsymbol{y} \in \texttt{dom} f$, so Lemma 11(c) gives $\{\boldsymbol{x}^U \sqcap^\circ \boldsymbol{y}^U, \boldsymbol{x}^U \sqcup^\circ \boldsymbol{y}^U\} = \{\boldsymbol{x}^U, \boldsymbol{y}^U\}$.

If $|\Delta(\boldsymbol{x}, \boldsymbol{y})| = 0$ then $\{\boldsymbol{x} \sqcap \boldsymbol{y}, \boldsymbol{x} \sqcup \boldsymbol{y}\} = \{\boldsymbol{x}, \boldsymbol{y}\}$ and so the claim holds trivially. Let us assume that $\Delta(\boldsymbol{x}, \boldsymbol{y}) = \{j\}$. We will write $\boldsymbol{x} = (\boldsymbol{x}^U, x_j, \boldsymbol{z})$ and $\boldsymbol{y} = (\boldsymbol{y}^U, y_j, \boldsymbol{z})$ where $\boldsymbol{z} = \boldsymbol{x}^{\overline{U} - \{j\}} = \boldsymbol{y}^{\overline{U} - \{j\}}$. Denote $\boldsymbol{z}^{01} = (\boldsymbol{x}^U, y_j, \boldsymbol{z})$ and $\boldsymbol{z}^{10} = (\boldsymbol{y}^U, x_j, \boldsymbol{z})$. Clearly, we have either $\{\boldsymbol{x} \sqcap \boldsymbol{y}, \boldsymbol{x} \sqcup \boldsymbol{y}\} = \{\boldsymbol{x}, \boldsymbol{y}\}$ or $\{\boldsymbol{x} \sqcap \boldsymbol{y}, \boldsymbol{x} \sqcup \boldsymbol{y}\} = \{\boldsymbol{z}^{01}, \boldsymbol{z}^{10}\}$. We can assume that the latter condition holds, otherwise (4) is a trivial equality. By Lemma 11(d) we have $(x_i, y_j), (y_i, x_j) \in \rho_{ij}$ for all $i \in U$, therefore $\boldsymbol{z}^{01}, \boldsymbol{x}^{10} \in \texttt{dom} f$. Two cases are possible:

<u>**Case 1**</u> $\{x_j, y_j\} \in M_j$, so $\sqcap_j^\circ, \sqcup_j^\circ$ are commutative on $\{x_j, y_j\}$. Thus, we must have either $\{\boldsymbol{x} \sqcap^\circ \boldsymbol{y}, \boldsymbol{x} \sqcup^\circ \boldsymbol{y}\} = \{\boldsymbol{z}^{01}, \boldsymbol{z}^{10}\}$ or $\{\boldsymbol{y} \sqcap^\circ \boldsymbol{x}, \boldsymbol{y} \sqcup^\circ \boldsymbol{x}\} = \{\boldsymbol{z}^{01}, \boldsymbol{z}^{10}\}$. Using the fact that $\langle \sqcap^\circ, \sqcup^\circ \rangle$ is a multimorphism of $f$, we get in each case the desired inequality:

$$f(\boldsymbol{z}^{01}) + f(\boldsymbol{z}^{01}) \leq f(\boldsymbol{x}) + f(\boldsymbol{y})$$

<u>**Case 2**</u> $\{x_j, y_j\} \in \overline{M}_j$. It can be checked that applying operations $\langle \texttt{Mj}^1, \texttt{Mj}^2, \texttt{Mn}^3 \rangle$ to $(\boldsymbol{x}, \boldsymbol{y}, \boldsymbol{z}^{01})$ gives $(\boldsymbol{z}^{01}, \boldsymbol{z}^{01}, \boldsymbol{z}^{10})$, therefore

$$f(\boldsymbol{z}^{01}) + f(\boldsymbol{z}^{01}) + f(\boldsymbol{z}^{10}) \leq f(\boldsymbol{x}) + f(\boldsymbol{y}) + f(\boldsymbol{z}^{01})$$

which is equivalent to (4).

□

**Proposition 16.** *If $\boldsymbol{x}, \boldsymbol{y} \in \texttt{dom} f$ and $\delta(\boldsymbol{x}, \boldsymbol{y}) = 1$ then either $\delta(\boldsymbol{x} \sqcup \boldsymbol{y}, \boldsymbol{y}) = 0$ or $\delta(\boldsymbol{x}, \boldsymbol{x} \sqcup \boldsymbol{y}) = 0$.*

*Proof.* Using the same argumentation as in the proof of Lemma 15 we conclude that $\{\boldsymbol{x}^U \sqcap \boldsymbol{y}^U, \boldsymbol{x}^U \sqcup \boldsymbol{y}^U\} = \{\boldsymbol{x}^U, \boldsymbol{y}^U\}$. If $\boldsymbol{x}^U \sqcup \boldsymbol{y}^U = \boldsymbol{x}^U$ then $\delta(\boldsymbol{x} \sqcup \boldsymbol{y}, \boldsymbol{y}) = 0$, and if $\boldsymbol{x}^U \sqcup \boldsymbol{y}^U = \boldsymbol{y}^U$ then $\delta(\boldsymbol{x}, \boldsymbol{x} \sqcup \boldsymbol{y}) = 0$. □

We now proceed with the induction argument. Suppose that $\Delta(\boldsymbol{x}, \boldsymbol{y}) \geq 2$. We can assume without loss of generality that $\delta(\boldsymbol{x}, \boldsymbol{y}) = 1$, otherwise the claim is trivial. Denote

$$
\begin{aligned}
X &= \{i \in \Delta(\boldsymbol{x}, \boldsymbol{y}) \mid x_i \sqcap y_i = x_i,\ x_i \sqcup y_i = y_i\} \\
Y &= \{i \in \Delta(\boldsymbol{x}, \boldsymbol{y}) \mid x_i \sqcap y_i = y_i,\ x_i \sqcup y_i = x_i\}
\end{aligned}
$$

We have $|X \cup Y| \geq 2$, so by Proposition 16 at least one of the two cases below holds:

<u>**Case 1**</u> $|X| \geq 2$ or $|X| = 1, \delta(\boldsymbol{x} \sqcup \boldsymbol{y}, \boldsymbol{y}) = 0$. It can be checked that $(\boldsymbol{x} \sqcup \boldsymbol{y}) \sqcap \boldsymbol{y} = \boldsymbol{y}$. Therefore, if we define $\boldsymbol{x}' = \boldsymbol{x} \sqcup \boldsymbol{y}$, $\boldsymbol{y}' = \boldsymbol{y}$ then the following identities hold:

$$
\boldsymbol{x} \sqcap \boldsymbol{y}' = \boldsymbol{x} \sqcap \boldsymbol{y} \qquad \boldsymbol{x} \sqcup \boldsymbol{y}' = \boldsymbol{x}' \qquad \boldsymbol{x}' \sqcap \boldsymbol{y} = \boldsymbol{y}' \qquad \boldsymbol{x}' \sqcup \boldsymbol{y} = \boldsymbol{x} \sqcup \boldsymbol{y} \tag{5}
$$

Let us select node $s \in X$ and modify $\boldsymbol{y}'$ by setting $y'_s = x_s$. (Note that we have $x'_s = x_s$.) It can be checked that (5) still holds. We have

- $(\boldsymbol{x}, \boldsymbol{y}') \prec (\boldsymbol{x}, \boldsymbol{y})$ since $\Delta(\boldsymbol{x}, \boldsymbol{y}') = \Delta(\boldsymbol{x}, \boldsymbol{y}) - \{s\}$, and
- $(\boldsymbol{x}', \boldsymbol{y}) \prec (\boldsymbol{x}, \boldsymbol{y})$ since $\Delta(\boldsymbol{x}', \boldsymbol{y}) = \Delta(\boldsymbol{x}, \boldsymbol{y}) - (X - \{s\})$; if $X - \{s\}$ is empty then $\delta(\boldsymbol{x}', \boldsymbol{y}) < \delta(\boldsymbol{x}, \boldsymbol{y})$.

Thus, by the induction hypothesis

$$
f(\boldsymbol{x} \sqcap \boldsymbol{y}) + f(\boldsymbol{x}') \leq f(\boldsymbol{x}) + f(\boldsymbol{y}') \tag{6}
$$

assuming that $\boldsymbol{y}' \in \texttt{dom} f$, and

$$
f(\boldsymbol{y}') + f(\boldsymbol{x} \sqcup \boldsymbol{y}) \leq f(\boldsymbol{x}') + f(\boldsymbol{y}) \tag{7}
$$

assuming that $\boldsymbol{x}' \in \texttt{dom} f$. If $\boldsymbol{y}' \in \texttt{dom} f$ then Inequality (6) implies that $\boldsymbol{x}' \in \texttt{dom} f$, and the claim then follows from summing (6) and (7). We now assume that $\boldsymbol{y}' \notin \texttt{dom} f$; Inequality (7) then implies that $\boldsymbol{x}' \notin \texttt{dom} f$.

Assume for simplicity of notation that $k$ corresponds to the first argument of $f$. Define instance $\hat{\mathcal{I}}$ with the set of nodes $\hat{V} = V - \{s\}$ and cost function

$$
g(\boldsymbol{z}) = \min_{a \in D_s} \{u(a) + f(a, \boldsymbol{z})\} \qquad \forall \boldsymbol{z} \in \hat{\mathcal{D}} \equiv \bigotimes_{i \in \hat{V}} D_i
$$

where $u(a)$ is the following unary cost function: $u(x_s) = 0$, $u(y_s) = C$ and $u(a) = 2C$ for $a \in D - \{x_s, y_s\}$. Here $C$ is a sufficiently large constant, namely $C > f(\boldsymbol{x}) + f(\boldsymbol{y})$. It is straightforward to check that unary relations $D_i, i \in \hat{V}$ and binary relations $\rho_{ij}, i, j \in \hat{V}, i \neq j$ are the unique arc- and path-consistent relations for $g$, i.e.

$$
\rho_i = \{x_i \mid \boldsymbol{x} \in \texttt{dom} g\} \quad \forall i \in \hat{V}\ , \qquad \rho_{ij} = \{(x_i, x_j) \mid \boldsymbol{x} \in \texttt{dom} g\} \quad \forall i, j \in \hat{V}, i \neq j
$$

This implies that set $U \subseteq \hat{V}$ and sets $A_i, B_i$ for $i \in U$ satisfy conditions (a-d) of Lemma 11 for instance $\hat{\mathcal{I}}$. Clearly, $\langle \sqcap^\circ, \sqcup^\circ \rangle$ is a multimorphism of $g$. Furthermore, if the modification in Stage 2 had been applied to instance $\hat{\mathcal{I}}$ and sets $U, A_i, B_i$ then it would give the same pair $\langle \sqcap, \sqcup \rangle$ that we obtained for $\mathcal{I}$. This reasoning shows that we can use the induction hypothesis for $\hat{\mathcal{I}}$: if $\boldsymbol{u}, \boldsymbol{v} \in \mathtt{dom}\, g$ and $(\boldsymbol{u}, \boldsymbol{v}) \prec (\boldsymbol{x}, \boldsymbol{y})$ then $g(\boldsymbol{u} \sqcap \boldsymbol{v}) + g(\boldsymbol{u} \sqcup \boldsymbol{v}) \leq g(\boldsymbol{u}) + g(\boldsymbol{v})$.

Let $\hat{\boldsymbol{x}}, \hat{\boldsymbol{y}}, \hat{\boldsymbol{x}}', \hat{\boldsymbol{y}}'$ be restrictions of respectively $\boldsymbol{x}, \boldsymbol{y}, \boldsymbol{x}', \boldsymbol{y}'$ to $\hat{V}$. We can write

$$\begin{aligned} g(\hat{\boldsymbol{y}}) = g(\hat{\boldsymbol{y}}') &= u(y_s) + f(y_s, \hat{\boldsymbol{y}}) = f(\boldsymbol{y}) + C \qquad \text{(since } (x_s, \hat{\boldsymbol{y}}) = \boldsymbol{y}' \notin \mathtt{dom}\, f) \\ g(\hat{\boldsymbol{x}}) &= f(x_s, \hat{\boldsymbol{x}}) = f(\boldsymbol{x}) \end{aligned}$$

By the induction hypothesis

$$g(\hat{\boldsymbol{x}} \sqcap \hat{\boldsymbol{y}}) + g(\hat{\boldsymbol{x}} \sqcup \hat{\boldsymbol{y}}) \leq g(\hat{\boldsymbol{x}}) + g(\hat{\boldsymbol{y}}) = f(\boldsymbol{x}) + f(\boldsymbol{y}) + C \tag{8}$$

We have $g(\hat{\boldsymbol{x}} \sqcup \hat{\boldsymbol{y}}) < 2C$, so we must have either $g(\hat{\boldsymbol{x}} \sqcup \hat{\boldsymbol{y}}) = f(x_s, \hat{\boldsymbol{x}} \sqcup \hat{\boldsymbol{y}})$ or $g(\hat{\boldsymbol{x}} \sqcup \hat{\boldsymbol{y}}) = f(y_s, \hat{\boldsymbol{x}} \sqcup \hat{\boldsymbol{y}}) + C = f(\boldsymbol{x} \sqcup \boldsymbol{y}) + C$. The former case is impossible since $(x_s, \hat{\boldsymbol{x}} \sqcup \hat{\boldsymbol{y}}) = \boldsymbol{x}' \notin \mathtt{dom}\, f$, so $g(\hat{\boldsymbol{x}} \sqcup \hat{\boldsymbol{y}}) = f(\boldsymbol{x} \sqcup \boldsymbol{y}) + C$. Combining it with (8) gives

$$g(\hat{\boldsymbol{x}} \sqcap \hat{\boldsymbol{y}}) + f(\boldsymbol{x} \sqcup \boldsymbol{y}) \leq f(\boldsymbol{x}) + f(\boldsymbol{y}) \tag{9}$$

This implies that $g(\hat{\boldsymbol{x}} \sqcap \hat{\boldsymbol{y}}) < C$, so we must have $g(\hat{\boldsymbol{x}} \sqcap \hat{\boldsymbol{y}}) = f(x_s, \hat{\boldsymbol{x}} \sqcap \hat{\boldsymbol{y}}) = f(\boldsymbol{x} \sqcap \boldsymbol{y})$. Thus, (9) is equivalent to (4).

**Case 2** $|Y| \geq 2$ or $|Y| = 1, \delta(\boldsymbol{x}, \boldsymbol{x} \sqcup \boldsymbol{y}) = 0$. It can be checked that $\boldsymbol{x} \sqcap (\boldsymbol{x} \sqcup \boldsymbol{y}) = \boldsymbol{x}$. Therefore, if we define $\boldsymbol{x}' = \boldsymbol{x}$, $\boldsymbol{y}' = \boldsymbol{x} \sqcup \boldsymbol{y}$ then the following identities hold:

$$\boldsymbol{x}' \sqcap \boldsymbol{y} = \boldsymbol{x} \sqcap \boldsymbol{y} \qquad \boldsymbol{x}' \sqcup \boldsymbol{y} = \boldsymbol{y}' \qquad \boldsymbol{x} \sqcap \boldsymbol{y}' = \boldsymbol{x}' \qquad \boldsymbol{x} \sqcup \boldsymbol{y}' = \boldsymbol{x} \sqcup \boldsymbol{y} \tag{10}$$

Let us select node $s \in Y$ and modify $\boldsymbol{x}'$ by setting $x'_s = y_s$. (Note that we have $y'_s = y_s$.) It can be checked that (10) still holds. We have $(\boldsymbol{x}', \boldsymbol{y}) \prec (\boldsymbol{x}, \boldsymbol{y})$ and $(\boldsymbol{x}, \boldsymbol{y}') \prec (\boldsymbol{x}, \boldsymbol{y})$ since $\Delta(\boldsymbol{x}', \boldsymbol{y}) = \Delta(\boldsymbol{x}, \boldsymbol{y}) - \{s\}$ and $\Delta(\boldsymbol{x}, \boldsymbol{y}') = \Delta(\boldsymbol{x}, \boldsymbol{y}) - (Y - \{s\})$, so by the induction hypothesis

$$f(\boldsymbol{x} \sqcap \boldsymbol{y}) + f(\boldsymbol{y}') \leq f(\boldsymbol{x}') + f(\boldsymbol{y}) \tag{11}$$

assuming that $\boldsymbol{x}' \in \mathtt{dom}\, f$, and

$$f(\boldsymbol{x}') + f(\boldsymbol{x} \sqcup \boldsymbol{y}) \leq f(\boldsymbol{x}) + f(\boldsymbol{y}') \tag{12}$$

assuming that $\boldsymbol{y}' \in \mathtt{dom}\, f$. If $\boldsymbol{x}' \in \mathtt{dom}\, f$ then Inequality (11) implies that $\boldsymbol{y}' \in \mathtt{dom}\, f$, and the claim then follows from summing (11) and (12). We now assume that $\boldsymbol{x}' \notin \mathtt{dom}\, f$; Inequality (12) then implies that $\boldsymbol{y}' \notin \mathtt{dom}\, f$.

Assume for simplicity of notation that $k$ corresponds to the first argument of $f$. Define instance $\hat{\mathcal{I}}$ with the set of nodes $\hat{V} = V - \{s\}$ and cost function

$$g(\boldsymbol{z}) = \min_{a \in D_s} \{u(a) + f(a, \boldsymbol{z})\} \qquad \forall \boldsymbol{z} \in \hat{\mathcal{D}} \equiv \bigotimes_{i \in \hat{V}} D_i$$

where $u(a)$ is the following unary term: $u(y_s) = 0$, $u(x_s) = C$ and $u(a) = 2C$ for $a \in D - \{x_s, y_s\}$. Here $C$ is a sufficiently large constant, namely $C > f(\boldsymbol{x}) + f(\boldsymbol{y})$.

Let $\hat{\boldsymbol{x}}, \hat{\boldsymbol{y}}, \hat{\boldsymbol{x}}', \hat{\boldsymbol{y}}'$ be restrictions of respectively $\boldsymbol{x}, \boldsymbol{y}, \boldsymbol{x}', \boldsymbol{y}'$ to $\hat{V}$. We can write

$$\begin{aligned} g(\hat{\boldsymbol{x}}) = g(\hat{\boldsymbol{x}}') &= u(x_s) + f(x_s, \hat{\boldsymbol{x}}) = f(\boldsymbol{x}) + C \qquad \text{(since } (y_s, \hat{\boldsymbol{x}}) = \boldsymbol{x}' \notin \mathtt{dom}\, f) \\ g(\hat{\boldsymbol{y}}) &= f(y_s, \hat{\boldsymbol{y}}) = f(\boldsymbol{y}) \end{aligned}$$

By the induction hypothesis

$$g(\hat{\boldsymbol{x}} \sqcap \hat{\boldsymbol{y}}) + g(\hat{\boldsymbol{x}} \sqcup \hat{\boldsymbol{y}}) \leq g(\hat{\boldsymbol{x}}) + g(\hat{\boldsymbol{y}}) = f(\boldsymbol{x}) + f(\boldsymbol{y}) + C \tag{13}$$

We have $g(\hat{\boldsymbol{x}} \sqcup \hat{\boldsymbol{y}}) < 2C$, so we must have either $g(\hat{\boldsymbol{x}} \sqcup \hat{\boldsymbol{y}}) = f(y_s, \hat{\boldsymbol{x}} \sqcup \hat{\boldsymbol{y}})$ or $g(\hat{\boldsymbol{x}} \sqcup \hat{\boldsymbol{y}}) = f(x_s, \hat{\boldsymbol{x}} \sqcup \hat{\boldsymbol{y}}) + C = f(\boldsymbol{x} \sqcup \boldsymbol{y}) + C$. The former case is impossible since $(y_s, \hat{\boldsymbol{x}} \sqcup \hat{\boldsymbol{y}}) = \boldsymbol{y}' \notin \mathtt{dom}\, f$, so $g(\hat{\boldsymbol{x}} \sqcup \hat{\boldsymbol{y}}) = f(\boldsymbol{x} \sqcup \boldsymbol{y}) + C$. Combining it with (13) gives

$$g(\hat{\boldsymbol{x}} \sqcap \hat{\boldsymbol{y}}) + f(\boldsymbol{x} \sqcup \boldsymbol{y}) \leq f(\boldsymbol{x}) + f(\boldsymbol{y}) \tag{14}$$

This implies that $g(\hat{\boldsymbol{x}} \sqcap \hat{\boldsymbol{y}}) < C$, so we must have $g(\hat{\boldsymbol{x}} \sqcap \hat{\boldsymbol{y}}) = f(y_s, \hat{\boldsymbol{x}} \sqcap \hat{\boldsymbol{y}}) = f(\boldsymbol{x} \sqcap \boldsymbol{y})$. Thus, (14) is equivalent to (4).